# Ultrafast optically induced magnetic state transition in 2D antiferromagnets

Shuo Li [a,b], Junjie He [c,d], Thomas Frauenheim [b,c,e]

[a] *Institute for Advanced Study, Chengdu University, Chengdu 610100, P. R. China*

[b] *Shenzhen JL Computational Science and Applied Research Institute, Shenzhen 518110, P. R. China*

[c] *Bremen Center for Computational Materials Science, University of Bremen, Bremen 28359, Germany*

[d] *Department of Physical and Macromolecular Chemistry, Faculty of Science, Charles University in Prague, Prague 12843, Czechia*

[e] *Beijing Computational Science Research Center, Beijing 100193, P. R. China*

*Email: shuoli.phd@gmail.com, junjie.he.phy@gmail.com, thomas.frauenheim@bccms.uni-bremen.de*

**Abstract:** Manipulating spin in antiferromagnetic (AFM) materials has great potential in AFM opto-spintronics. Laser pulses can induce a transient ferromagnetic (FM) state in AFM metallic systems, but have never been proven in two-dimensional (2D) AFM semiconductors and related van der Waals (vdW) heterostructures. Here, using 2D vdW heterostructures of FM $MnS_2$ and AFM MXenes as prototypes, we investigated optically induced interlayer spin transfer dynamics based on the real-time time-dependent density functional theory (rt-TDDFT). We observed that laser pulses induce significant spin injection and the interfacial atom-mediated spin transfer from $MnS_2$ to $Cr_2CCl_2$. In particular, we first demonstrated the transient FM state in semiconducting AFM/FM heterostructures during photoexcited processes. Because the proximity magnetism breaks the magnetic symmetry of $Cr_2CCl_2$ in heterostructures. Our results provide the microscopic understanding for optically controlled interlayer spin dynamics in 2D magnetic heterostructures and open a new way to manipulate magnetic orders in ultrafast opto-spintronics.

**Keywords**

spin injection, transient ferromagnetic state, heterostructures, photoinduced spin dynamics, rt-TDDFT

# 1. Introduction

Two-dimensional (2D) magnetic materials have great potential for the application in spintronics, valleytronics, and electromagnetics.[1, 2] The intrinsic 2D magnetism was experimentally achieved in thin layers of $CrI_3$,[3] $Cr_2Ge_2Te_6$,[4] $Fe_3GeTe_2$,[5] and $MnSe_2$,[6] in which the long range magnetic ordering can be stabilized by magnetic anisotropy with an excitation gap opening. Moreover, these 2D magnetisms are tunable with the layer thickness,[3, 7] electric field,[8, 9] strain,[10, 11] and light [12] for applications in information technology and quantum computing. In particular, quantum properties of 2D magnetic materials can be tailored by van der Waals (vdW) interfacial engineering, which can introduce new functionalities in 2D spintronics devices.[13] Moreover, the magnetic proximity effect (MPE) can introduce novel physical properties in heterostructures, such as spin injection, tunneling spin-valve, valley polarization, and quantum anomalous Hall states.[1, 13-16] Large scale 2D magnetic transition metal dichalcogenides (TMDs) can experimentally prepared.[17, 18] In particular, monolayer $MnSe_2$ [6] and $VSe_2$ [19] with Curie temperature ($T_c$) close to and even above room temperature has been demonstrated. Moreover, theoretical studies[20, 21] also showed that 2D $MnS_2$ and $MnSe_2$ sheets are ideal ferromagnetic (FM) semiconductors with long-range magnetic ordering high magnetic moments high $T_c$ and tunable magnetic anisotropy. The results suggest that Manganese-based TMDs have great potential as 2D spintronic devices.

Among these 2D magnets, 2D antiferromagnetic (AFM) materials are particularly interesting,[22] because the absence of a primary macroscopic magnetization makes spin manipulation in antiferromagnets inherently faster than in ferromagnets.[23] 2D AFM materials are permitted in each magnetic symmetry group and also offer much greater structural flexibility than FM materials because there are many ways arranging magnetic moments to achieve a zero net moment.[23, 24] However, 2D AFM materials are relatively less studied and far less utilized. Because controlling spins in antiferromagnets require high magnetic fields and AFM order cannot be detected with conventional magnetometry. MXenes (2D transition metal carbides or nitrides) are experimentally exfoliated from the MAX phases in acid solution, such as HF.[25] The FM behaviour of pristine $Cr_2C$ can become semiconducting antiferromagnetic (AFM) states with F, Cl and OH functionalization. Moreover, asymmetrical functionalization can induce different magnetic properties of $Cr_2C$ MXenes, for example $Cr_2COF$ MXenes with a strong spin-valley coupling[26] and the bipolar AFM semiconducting $Cr_2CFCl$ MXene.[27] Therefore, $Cr_2C$ MXenes provide an excellent platform to study the tunable magnetism and spin transfer in 2D AFM system.

In opto-spintronics, by using ultrashort laser pulses in the ranging from attoseconds to femtoseconds, the spin of magnetic materials can be manipulated for reaching the fastest information recording and processing and least-dissipative power.[28] The ultrafast spin injection and the ultrafast demagnetization are highly relevant for advancing ultrafast spintronics.[29-32] In this context, based on the real-time time-density functional theory (rt-TDDFT) simulation, a laser can redistribute spins between different magnetic sublattices by optically induced spin transfer (OISTR) effects.[33] OISTR represents the fastest mechanism to control the spin of matter results from the direct interaction between the spin of the material and the light field itself and, OISTR have been experimentally confirmed in magnetic metal multilayers and in Heusler compounds. Later, He et al. identified an interfacial atom-mediated spin transfer (IAMSTR) pathway in heterostructures.[34]

Many investigations focused on ultrafast optical quenching of magnetic order in FM materials.[35-41] On the contrary, reports on AFM materials are scarce due to the zero net magnetic moment. Previous works[42] showed that manipulating AFM order is faster than FM and provide the advanced perspective for the cooperative utilization of FM and AFM components in opto-spintronics.[22, 23] Therefore, magnetic materials can switch between AFM and FM order on ultrafast timescales, such as FeRh exciting with fs laser pulses.[43, 44] Radu et al. reported a transient FM state in ferromagnetic material GdFeCo at the picosecond timescale.[31] Recently, Golias et al. an ultrafast optically induced FM alignment of AFM Mn in Co/Mn multilayers.[45] These investigations reported on the transient FM state in the excitation of metallic systems with asymmetrical AFM alignment. However, the switching of magnetic ordering (e.g. transient FM state) in 2D vdW heterostructures are not clear. Therefore, studying photoinduced spin dynamics and the underlying mechanism of interlayered spin transfer and the transient FM state in 2D magnets and vdW heterostructures will promote the development of AFM opto-spintronics.

In this work, using $Cr_2CCl_2$-$MnS_2$ vdW heterostructures as prototypes, we theoretically investigated the photoinduced spin dynamics in representative AFM-FM $Cr_2CCl_2$-$MnS_2$ vdW heterostructures, based on rt-TDDFT. Our results show that laser pulses induce a transient FM state of AFM $Cr_2CCl_2$ in a few femtoseconds. The microscopic mechanism for OISTR and IAMSTR in AFM-FM heterostructures has also been unraveled in this study.

## 2. Results and Discussion

Optimized lattice parameters of $MnS_2$ and $Cr_2CCl_2$ are 3.32 and 3.13 Å at the PBE level, in which the lattice mismatch is 5.7%. Different stacking configurations of $Cr_2CCl_2$-$MnS_2$ heterostructures have been considered (Figure S1). To estimate the structural stabilities of the heterostructures under different stacking patterns, we calculated the binding energies, which are defined as $E_b = E_{MXene-TMD} - E_{MXene} - E_{TMD}$. Here, $E_{MXene-TMD}$ is the total energy of the heterostructure, and $E_{MXene}$ and $E_{TMD}$ are the total energy of the isolated MXene and TMD layers, respectively. $E_b$ of $Cr_2CCl_2$ deposited on $MnS_2$ is -0.23 eV per unit cell, thus reflecting weaker interlayer interactions. After structural optimization, we identified the most favourable stacking structure of $Cr_2CCl_2$-$MnS_2$ heterostructures (Figure 1). The equilibrium distance between $MnS_2$ and $Cr_2CCl_2$ is 3.171 Å, which is relatively stronger interlayered interactions.

Herein, we consider two magnetic states for the $MnS_2$ substrate: interlayer AFM state and interlayer FM state in $Cr_2CCl_2$-$MnS_2$ heterostructures, in which the energy difference for these states is 0.6 meV. Moreover, the achievement of non-volatile reversal of magnetization of Mn element can be desired using an electric field.[9, 46, 47] The previous work showed $Cr_2CCl_2$ has a high magnetic transition temperature, which means a strong AFM order.[26] The spin-polarized calculations show that antiparallel spin polarizations in AFM $Cr_2CCl_2$ are broken, as expected considering MPE from the magnetic substrate. The MPE of $MnS_2$ results in a difference of local magnetic moments of 0.002 $\mu_B$ between $Cr_1$ and $Cr_2$. Moreover, the projected band structures of the $MnS_2$, $Cr_2CCl_2$ and their heterostructures are shown in Figure S2, where $MnS_2$ and $Cr_2CCl_2$ are FM and AFM semiconductors respectively, in agreement with previous works.[20, 26] When the AFM $Cr_2CCl_2$ is deposited onto $MnS_2$, the MPE from the $MnS_2$ substrate will slightly change the electronic structure of $Cr_2CCl_2$.

In Figures 1a and 1b, the switching of magnetic orders from AFM to transient FM in $Cr_2CCl_2$-$MnS_2$ heterostructures. The microscopic mechanism of ultrafast switching of magnetic orders is dominated by spin-selective charge transfer from one magnetic sublattice to another. The underlying principle of the OISTR mechanism are shown in Figures 1c and 1d. The laser (above a threshold value) can induces coherent electrons transfer between atoms in $Cr_2CCl_2$-$MnS_2$ heterostructures. In the AFM $Cr_2CCl_2$, majority states from $Cr_1$ are transferred to the minority of $Cr_2$ (and vice versa). When the asymmetry is present in the interface between Mn and $Cr_2$, a spin charge filling imbalance due to MPE is introduced and a transient FM alignment emerges (Figure 1c). Moreover, when the antiparallel spin occurs between Mn and $Cr_1$, a continuous filling of spin charges between magnetic sublattices will introduces the phase transfer from AFM to FM states (Figure 1d). Therefore, tuning magnetic orders of FM substrate can manipulate the transient FM alignment from AFM materials.

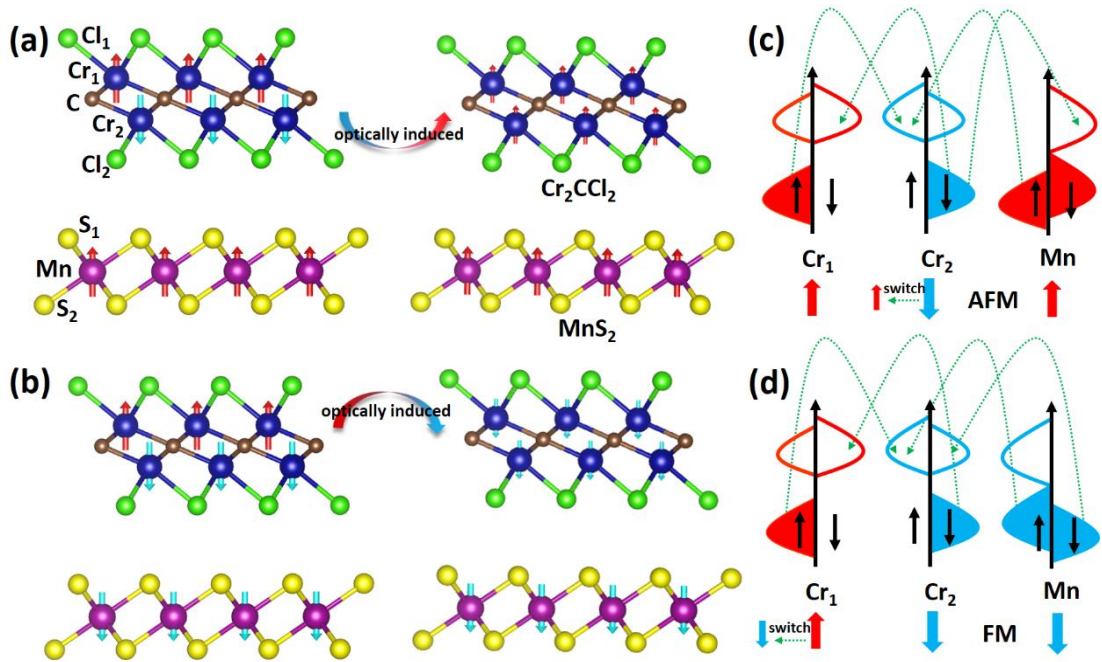

**Figure 1.** Schematics of magnetic configurations of (a) interlayer AFM state and (b)

interlayer FM state in $Cr_2CCl_2$-$MnS_2$ heterostructures. Atom legend: purple, Mn; blue, Cr; brown, C yellow, S; and green, Cl. The underlying principle of the OISTR mechanism for switching magnetic orders of (a) interlayer AFM state and (b) interlayer FM state in $Cr_2CCl_2$-$MnS_2$ heterostructures. Red and blue arrows are spin-up and spin-down, respectively.

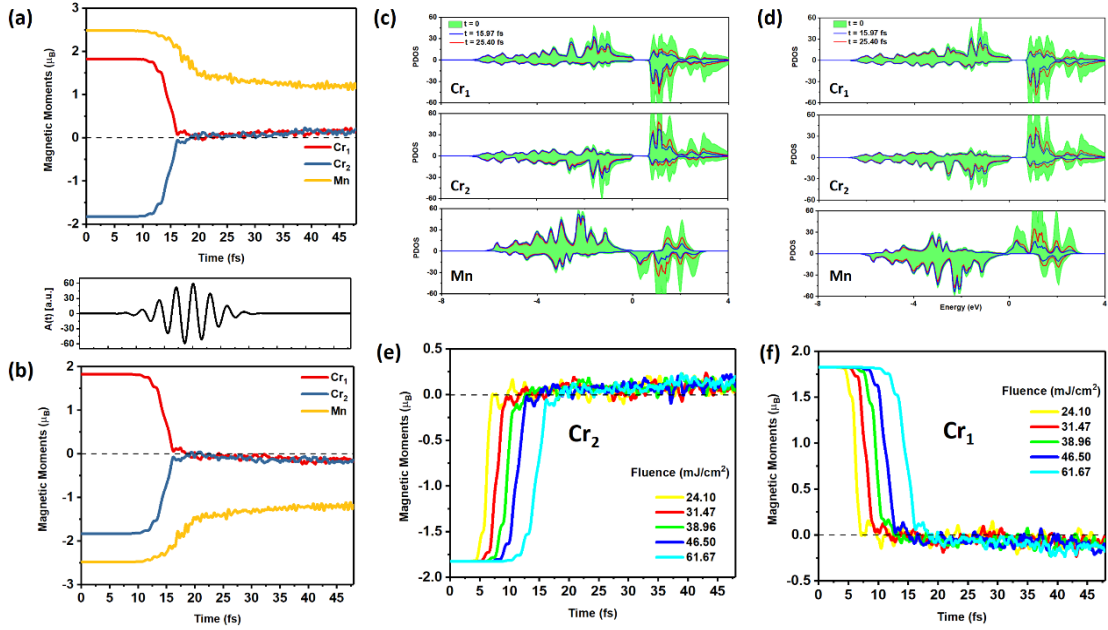

**Figure 2.** Switching of magnetic order of (a) interlayer AFM state and (b) interlayer FM state in $Cr_2CCl_2$-$MnS_2$ heterostructures. Projected density of states (DOS) of $Cr_1$, $Cr_2$ and Mn atom for (c) interlayer AFM state and (d) interlayer FM state in $Cr_2CCl_2$-$MnS_2$ heterostructures. Spin dynamics of local magnetic moment of (e) $Cr_2$ atom for interlayer AFM state and (f) $Cr_1$ atom for interlayer FM state under the influence of laser pulses of fluences.

In Figures 2a and 2b, we clearly show the transition from AFM to FM alignment of Cr layers after the arrival of a strong laser[48] (F = 61.67 mJ/cm$^2$, frequency=1.36 eV). The onset of the FM states shows that the laser reaches its half maximum and peaks simultaneously with its vector potential, while Mn shows a slower demagnetization. The underlying mechanism for the transient FM alignment is revealed in Figures 2c and 2d, where the unoccupied minority spin density of states (DOS) acts as a sink for excited majority spin electrons from the neighbouring Cr layer. The spin swapping between Cr neighbours facilitated by their AFM coupling, as well as the higher unoccupied state filling of the atoms at the interface ($Cr_2$) from the AFM-coupled reservoir of Mn majority electrons drive the transient FM state in Cr sublattices. Moreover, we found that Hubbub U values (see Supporting Information) for Cr and Mn atoms have no significant effect on the transient FM state, as shown in Figure S3.

The switching of magnetic order from AFM to FM layers can be understood as spin-selective charge transfer from one magnetic sublattice to others in 2D heterostructures, whereby the photoexcited spin of Mn first transfers to intralayered S atoms and then hops to the $Cr_2CCl_2$ MXene. The process of photoexcited spin transfer from Mn to intralayered S atoms is governed by optically induced intersite spin transfer (OISTR) physics. Moreover, the mechanism of OISTR depends on purely optical excitation of charge between magnetic sublattices. In general, shorter laser pulses could produce correspondingly faster changes in magnetization. Therefore, we simulated different fluences of the laser depending on a full width at half maximum (fwhm, from 3.63 fs to 9.68 fs) (Figures 2e and 2f). The results indicated that the optical switching of magnetic order occurs in two cases. The spin-selective exchange of charge between majority and minority DOS of magnetic sublattices. Moreover, we see that switching of the magnetic order occurs at earlier times, between 10 and 20 fs.

The spin injection, demagnetization and the switching of the magnetic order in heterostructures is sensitive to the selected fluence of the laser, thus indicating that this laser parameter can also be used to optimize and control the spin injection process. The change in fluence does not significantly affect the switching of the magnetic order, thereby indicating that our results are robust.

The magnetization dynamics of S and Cl atoms also provide more details (Figures 3a and 3b). S atoms gain spin rapidly in the time window [10, 18] fs. Concurrently, the magnetic order of $Cl_2$ atoms show a switching at ≈ 18 fs. The curve of Cl moment starts to increase after the S moment almost saturated at 18 fs. The photoexcited spin transfer through interlayer hopping (from S to Cl) is slower than intralayer spin transfer (from Mn to S). A slight increase in magnetic moments of C atom could be ignored. The timeresolved occupation dynamics show that the photoinduced spin-selective charge flowed across the vdW gap, resulting in the switching of magnetic order of MXenes. In addition, interlayer hopping was also slower than intralayer charge transfer. To gain further insights into the switching processes between AFM and to FM, we examined the time evolution of magnetization density in real space (Figures 3c and 3d).

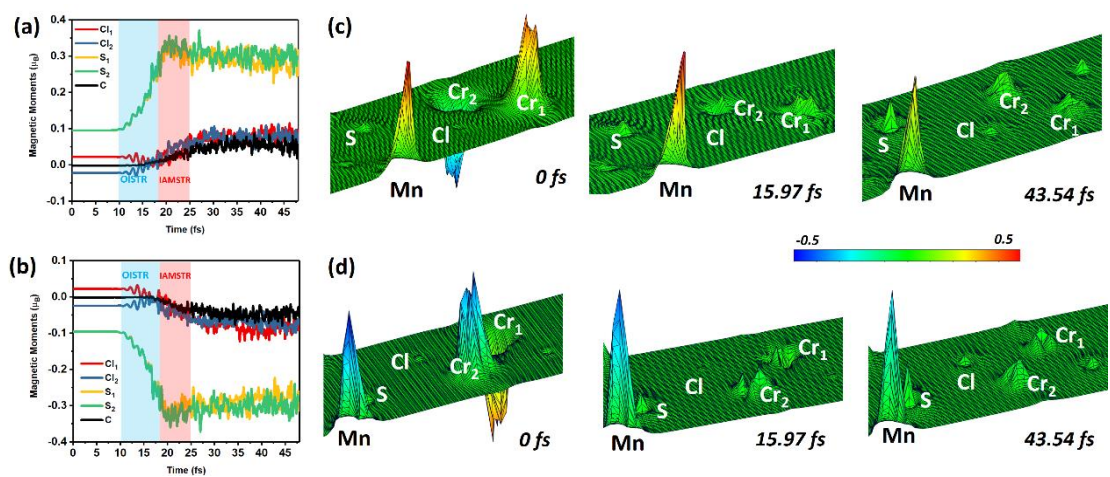

**Figure 3.** Time evolution of the local magnetic moment of S, Cl and Cr atoms for (a) interlayer AFM state and (b) interlayer FM state in $Cr_2CCl_2$-$MnS_2$ heterostructures. The time windows for OISTR and IAMSTR are marked in panels (a) and (b), respectively. Snapshots of the magnetization density at t = 0, 15.97, and 43.54 fs for (c) interlayer AFM state and (d) interlayer FM state of $Cr_2CCl_2$-$MnS_2$ heterostructures. The unit of the color scale is the intensity of magnetic moment.

The switching processes between AFM and to FM (Figure 3) can be divided into two time scales: (1) the time window [10, 18] fs, during which period the OISTR mechanism dominates the direct spin transfer from Mn to S and Cr demagnetization, (2) at the time window [18, 25] fs, when interfacial S atom-mediated spin transfer dynamics and injection to Cl layer occur, which is IAMSTR mechanism. The results indicate that unoccupied states play a key role in the spin transfer dynamics and in the demagnetization of 2D vdW heterostructures.

Previous reports on spin-orbital coupling (SOC) plays a key role in spin transfer and demagnetization. Hence, we simulated spin dynamics for $Cr_2CCl_2$-$MnS_2$ heterostructures with SOC, as shown in Figure S4. When SOC switches off, demagnetization can only occur due to spin transfer.[34, 40] When SOC switches on, the spin flip of photoexcited processes in $Cr_2CCl_2$-$MnS_2$ occurs because the total moment is no longer a good quantum number. The demagnetization process[29, 34] in $Cr_2CCl_2$-$MnS_2$ heterostructures can also be divided into three separate time scales: (1) purely spin current-induced demagnetization (t < 25 fs), (2) mixed contribution from spin flips and spin currents (25 < t < 37 fs), and (3) purely spin flip-dominated demagnetization (t > 37). While, SOC has no significant effect on spin injection of $Cr_2CCl_2$.

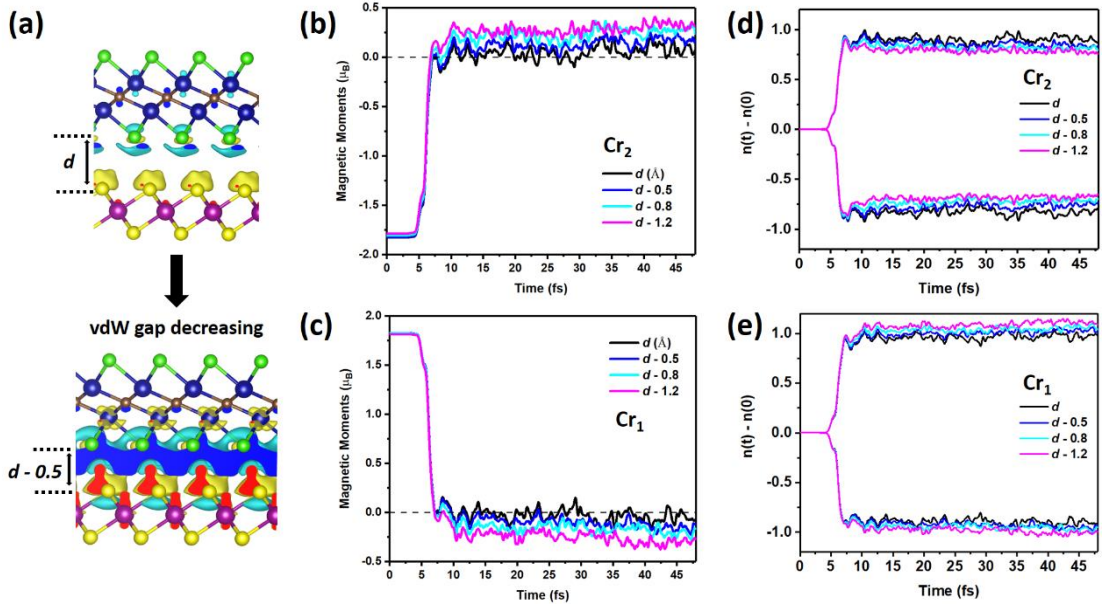

**Figure 4.** (a) Schematic diagram of the charge density difference of $Cr_2CCl_2$-$MnS_2$ heterostructure. The blue (yellow) regions represent the charge depletion (accumulation) with an isosurface value of 0.0001 e/Å$^3$. Spin dynamics of local magnetic moment of (b) $Cr_2$ atom for interlayer AFM state and (c) $Cr_1$ atom for interlayer FM state under the influence of vdw gaps. Number of majority and minority electrons, as a function of time (in fs), for (d) $Cr_2$ atom for interlayer AFM state and (e) $Cr_1$ atom for interlayer FM state in $Cr_2CCl_2$-$MnS_2$ heterostructures.

Based on the analysis of the mechanism of spin transfer dynamics, the switching of magnetic order occurs in the stage of IAMSTR, because of the spin-selective charge transfer in 2D vdW heterostructures through MPE. Therefore, decreasing vdW gap will make MPE stronger and make transient ferromagnetism more pronounced. The initial vdW gap and charge density differences (CDDs) of $Cr_2CCl_2$-$MnS_2$ heterostructures are shown in Figure 4a. The remarkable charge redistribution occurs at the interface of $Cr_2CCl_2$-$MnS_2$ heterostructures after decreasing the vdW gap of 0.5 Å, resulting in a strong MPE in $Cr_2CCl_2$-$MnS_2$ heterostructures.

We now turn to explore the light response on magnetism for $Cr_2CCl_2$-$MnS_2$ heterostructures with respect to the changing vdW gap. The magnetization dynamics of Cr atoms in $Cr_2CCl_2$ MXenes under the influence of vdW gaps are shown in Figures 4b and 4c. The transient FM moment of Cr atoms increases as vdW gap decreases. Moreover, the variation of the number of majority and minority electrons in the magnetic Cr layers with the changing vdW gaps, as a function of time, is shown in Figures 4d and 4e. The decrease (or increase) in the number of majority (or minority) electrons correspond to the demagnetization of the $Cr_1$ (or $Cr_2$) moment. Conversely, the gain of majority spins and loss of minority spins in Cr atoms accounts for their enhanced magnetic moment. Moreover, after decreasing the vdW gap, the interlayer coupling will be no longer a weak vdW interaction, but a strong Coulomb interaction, which may be similar to the system of metallic multilayer.[45] The results indicate that the photoinduced spin-selective charge flow results in demagnetization, spin injections and transient FM in $Cr_2CCl_2$-$MnS_2$ heterostructures and these processes involve spin currents of both minority and majority electrons.

## 3. Conclusions

Summarily, we investigated the ground-states properties and photoinduced spin dynamics of AFM-FM $Cr_2CCl_2$-$MnS_2$ heterostructures by performed DFT and real-time TDDFT calculations. Our DFT calculations showed that the proximity magnetism breaks the magnetic symmetry in AFM materials by inducing the magnetic substrate $MnS_2$. After optical excitation, laser pulses induce the significant spin injection from $MnS_2$ to $Cr_2CCl_2$ within a few femtoseconds. Moreover, we showed the interfacial atom-mediated spin transfer pathway in 2D heterostructures. Most importantly, the transient FM in 2D AFM materials with FM substrates can be induced by proximity magnetism. The amplitude of the transient FM of this 2D heterostructure is governed by the density of unoccupied states in AFM material. Overall, our results not only demonstrate that the transient FM in 2D AFM materials can be photoinduced by ultrafast lasers, but also open up new opportunities for proximity magnetism and spin transfer in 2D magnetic vdW heterostructures by using optical approaches.

## 4. Experimental Section

Calculations at ground state were performed at the density functional theory (DFT) level using the Vienna ab initio simulation package (VASP),[49, 50] using the Perdew–Burke–Ernzerhof (PBE) exchange–correlation functional.[51] The semi-empirical DFT-D3 method[52] was used to treat van der Waals (vdW) interactions. The Brillouin zone was represented by a Monkhorst–Pack Γ k-point mesh of 11 × 11 × 1 for structure relaxation at PBE level. To correctly determine the electronic and magnetic properties in strongly correlated systems containing transition metals, 15 × 15 × 1 k-points in Brillouin zone for unitcell, were represented at PBE+U level.[53] A correction of U = 3 eV for Cr and Mn is employed based on the relevant previous reports.[27, 54] An energy cut-off of 500 eV was used to determine the self-consistent charge density for the plane wave basis sets. The structures were fully optimized until the maximum Hellmann–Feynman force on atoms was lower than 0.01 eV Å$^{-1}$ and the total energy variation was lower than 1.0 ×10$^{-6}$ eV. A vacuum of approximately 15 Å was added in the perpendicular direction to the slab to model heterostructures.

To identify the spin dynamics of Cr$_2$CCl$_2$-MnS$_2$ vdW heterostructures under the influence of ultrafast laser pulses, we performed real-time time-dependent density functional theory (rt-TDDFT) calculations. The time evolving state functions ($\psi$) were calculated by solving the time dependent Kohn-Sham (KS) equation as follows:

$$i\frac{\partial \psi_j(\mathbf{r},t)}{\partial t} = \left[\frac{1}{2}\left(-i\nabla + \frac{1}{c}\mathbf{A}_{\text{ext}}(t)\right)^2 + v_s(\mathbf{r},t) + \frac{1}{2c}\sigma \cdot \mathbf{B}_s(\mathbf{r},t) + \frac{1}{4c^2}\sigma \cdot (\nabla v_s(\mathbf{r},t) \times -i\nabla)\right]\psi_j(\mathbf{r},t) \quad (1)$$

where $\mathbf{A}_{\text{ext}}(t)$ and σ represent vector potential and Pauli matrices. The KS effective potential $v_s(\mathbf{r},t) = v_{ext}(\mathbf{r},t) + v_H(\mathbf{r},t) + v_{xc}(\mathbf{r},t)$ can be decomposed into the external potential $v_{ext}$, the classical Hartree potential $v_H$, and the exchange-correlation (XC) potential $v_{xc}$, respectively. The KS magnetic field can be written as $B_s(r, t) = B_{ext}(r, t) + B_{xc}(r, t)$, where $B_{ext}$ and $B_{xc}$ represent the magnetic field of the applied laser pulse with an additional magnetic field and XC magnetic field, respectively. The last term in Eq. (1) stands for spin-orbital coupling effect.

In our rt-TDDFT simulations, we only considered the spin polarized electron dynamics based on the Born-Oppenheimer approximation to a time scale of t < 50 fs. This time scale is completely dominated by direct optical manipulation without the rotation of atomic magnetic moments.[55] The electron-phonon coupling effects are not taken into account. Photoinduced dynamics calculations were made using a fully non-collinear version of rt-TDDFT and a full-potential augmented plane-wave ELK code.[56] A regular mesh in a k-space of $6 \times 6 \times 1$, a smearing width of 0.027 eV, and a time step of $\Delta t = 0.1$ a.u. were used to simulate excited dynamics. The laser pulses that were used in the present study were linearly polarized (in-plane polarization) at a selected frequency. All calculations were performed using adiabatic local spin density approximations (ALSDA),[57] with U = 3 eV for Mn and Cr atoms.

**Supporting Information**

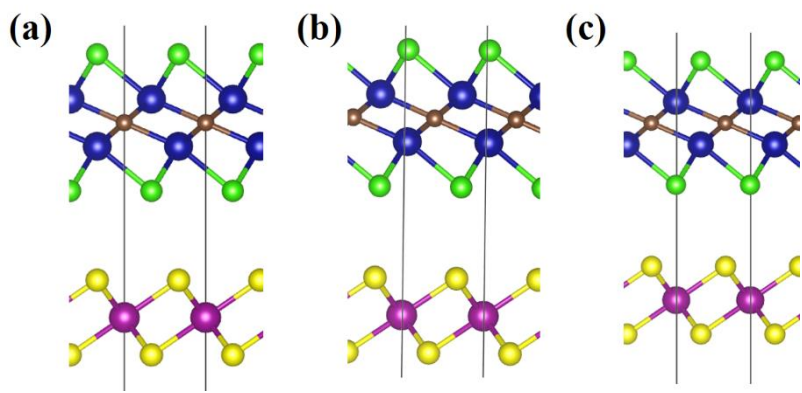

**Figure S1.** Side view of stacking configurations in models 1 (a), 2 (b) and 3 (c) of $Cr_2CCl_2$-$MnS_2$ vdW heterostructures. Model 1 is the most favourable stacking structure. Atom legend: purple, Mn; blue, Cr; brown, C yellow, S; and green, Cl.

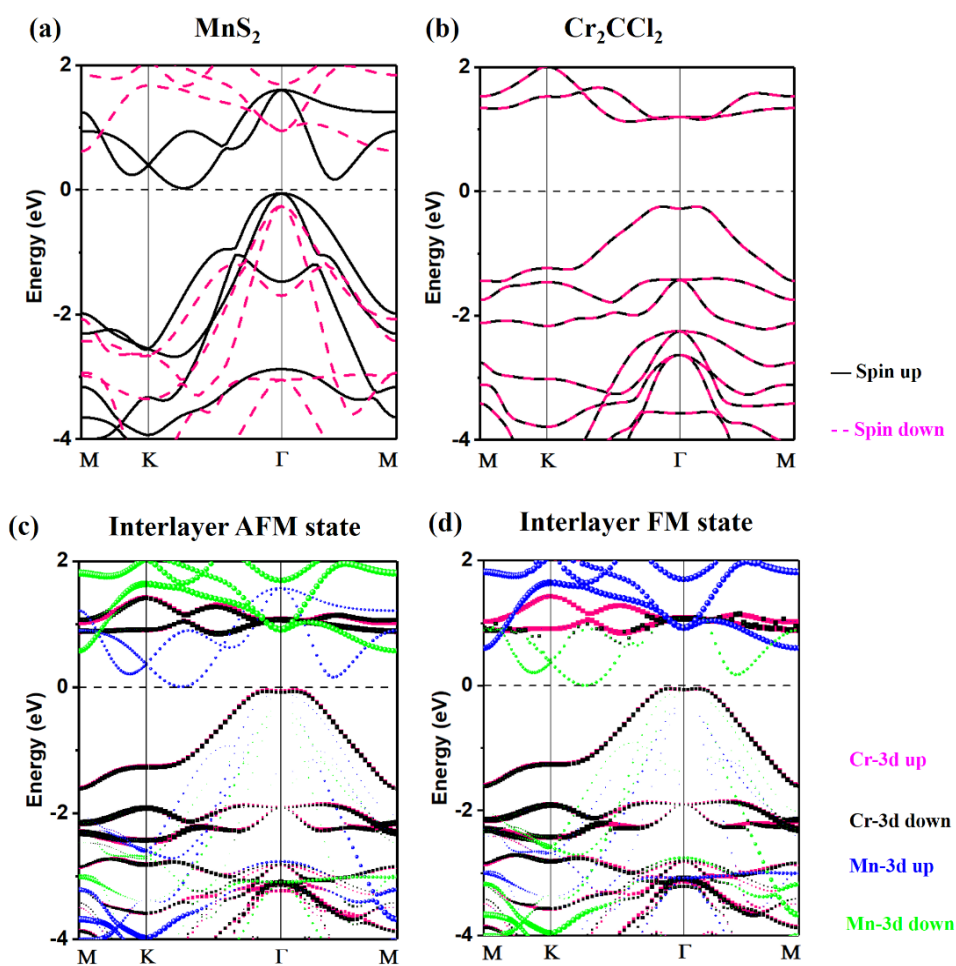

**Figure S2.** Band structures for (a) $MnS_2$ and (b) $Cr_2CCl_2$ monolayer and projected band structures for (c) interlayer AFM state and (d) interlayer FM state of $Cr_2CCl_2$-$MnS_2$

heterostructures.

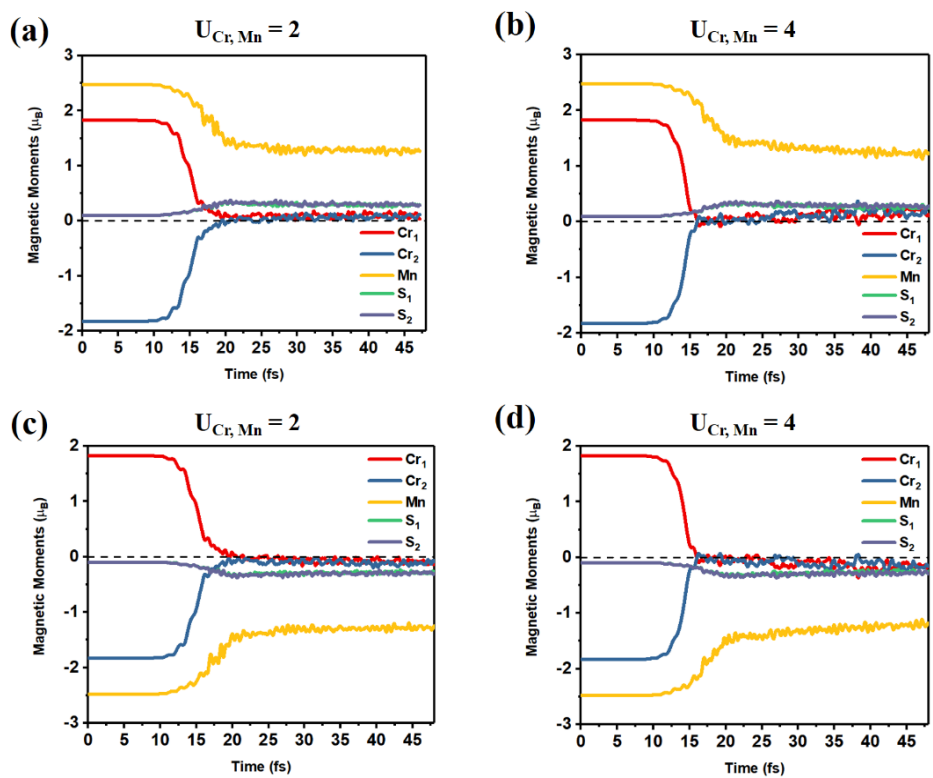

**Figure S3.** Switching of magnetic order of interlayer AFM state under U = 2 (a) and U = 4 (b) and interlayer FM state U = 2 (c) and U = 4 (d) in $Cr_2CCl_2$-$MnS_2$ heterostructures.

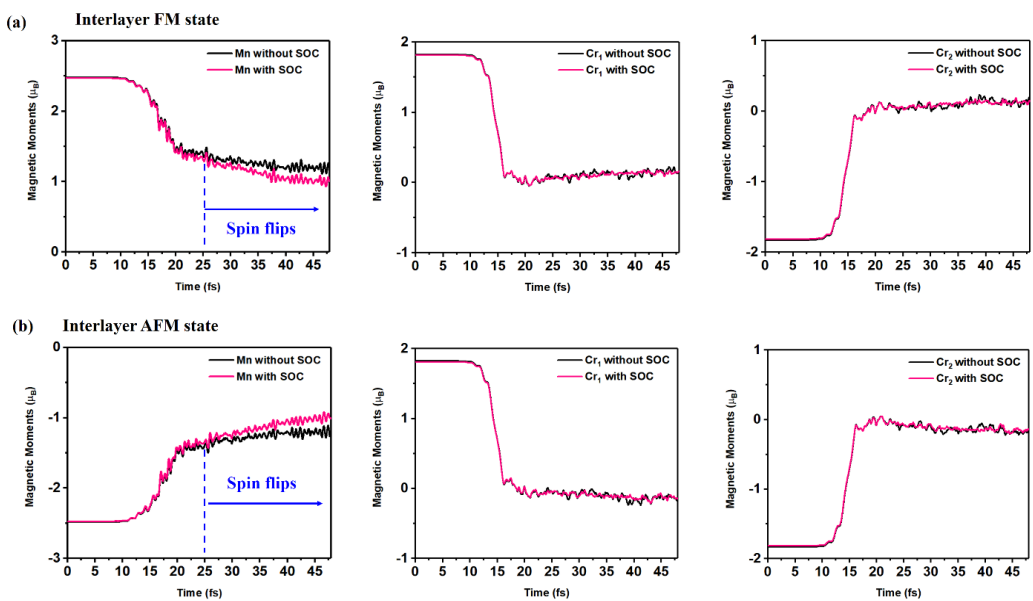

**Figure S4.** Changes in the magnetic moment of Mn and Cr atoms for (a) interlayer AFM state and (b) interlayer FM state, as a function of time (in fs), calculated without (black lines) SOC and with SOC (pink lines).